\documentclass{sig-alternate}

\usepackage{amsmath}
\usepackage{graphicx}
\usepackage{subfigure}
\usepackage{epsfig}
\usepackage{bm}

\newcommand{\commentout}[1]{}
\newcommand{\parahead}[1]{\vspace{0.07in}\noindent {\bf #1:}}

\newcommand{\myItemizeBegin}{
 \begin{list}{$\bullet$}{
    \setlength{\itemsep}{0pt}
    \setlength{\parsep}{3pt}
    \setlength{\topsep}{3pt}
    \setlength{\partopsep}{0pt}
    \setlength{\leftmargin}{2.0em}
    \setlength{\labelwidth}{1.5em}
    \setlength{\labelsep}{0.5em}
}}
\newcommand{\myItemizeEnd}{\end{list}}

\begin{document}


\title{Multi-Faceted Ranking of News Articles using Post-Read Actions}


\numberofauthors{3}
\author{
\alignauthor
Deepak Agarwal\\
       \affaddr{Yahoo! Research}\\
       \affaddr{dagarwal@yahoo-inc.com}
\alignauthor
Bee-Chung Chen\\
       \affaddr{Yahoo! Research}\\
       \affaddr{beechun@yahoo-inc.com}
\alignauthor
Xuanhui Wang\\
       \affaddr{Yahoo! Labs}\\
       \affaddr{xhwang@yahoo-inc.com}
}

\maketitle
\begin{abstract}

Personalized article recommendation is important to 
improve user engagement on news sites. Existing work
quantifies engagement primarily through click rates. 
We argue that quality of recommendations can be improved by incorporating different types of
``post-read'' engagement signals like sharing, commenting, printing and e-mailing
article links. More specifically, we propose a multi-faceted ranking problem for 
recommending news articles where each facet corresponds to a ranking problem to maximize actions of a
post-read action type. The key technical challenge is to estimate the rates of
post-read action types by mitigating the impact of enormous data sparsity, we do
so through several variations of factor models.
To exploit correlations among post-read action types we also introduce a novel variant called locally augmented tensor (LAT)
model. Through data obtained from a major news site in the US, we show that factor models
significantly outperform a few baseline IR models and the LAT model significantly outperforms
several other variations of factor models. Our findings show that it is possible to incorporate post-read
signals that are commonly available on online news sites to improve quality of recommendations.

\commentout{
To our surprise, we find lack of
correlation between click rates and post-read actions, suggesting
insufficiency of click-rate-based news recommendation.  Further, we
find anti-correlation between users' sharing and reading behavior for
some news categories, showing two distinct sides (public vs. private)
of users' news consumption behavior.  Based on these findings, we
propose a multi-faceted ranking problem for news articles, where each
facet corresponds to a post-read action type.  Through extensive
experiments using data collected from an online news site, we show
that factorization-based methods significantly outperform a cosine
similarity-based baseline method.  We also show that a newly proposed
Locally Augmented Tensor (LAT) model is able to exploit correlations
among different post-read engagement types and achieve the best
ranking performance among several other strong baselines.
}
\end{abstract}

\section{Introduction}
\label{sec:intro}
Publishing links to news articles has become
important to facilitate information discovery on the Web. Users
visiting a news website do not have a specific objective in mind
and simply want to be informed about news topics that are important
to them, or learn about topics that are of interest. Quality of
recommended links is crucial to ensure good user engagement in both
short and long terms. But the explicit signals about what the user
truly wishes to see is typically weak. Thus, it is important to
consider a broad array of complementary indicators of users'
interests --- novel techniques which can effectively leverage these weak signals are desired.

The primary indictor of user engagement used in most existing work
is the observed click-through rate or CTR, i.e., the probability
that a user would click an article when a link to the article is
displayed, and articles are usually ranked to optimize for
it~\cite{DasWWW07,coke:nips08,Li:www10}. We argue that merely using
CTR to rank news articles is not sufficient since user interaction
with online news has become multi-faceted. Users no longer simply
click on news links and read articles --- as shown in
Figure~\ref{fig:example}, they can share it with friends, tweet
about it, write and read comments, rate other users' comments,
email the link to friends and themselves, print the article to read
it thoroughly offline, and so on. These different types of ``post-read'' actions
are indicators of deep user engagement from different facets and
can provide additional signals for news recommendations.  We will use {\em facet} and {\em post-read action type} interchangeably.  For
example, news articles can be ranked for individual facets based on
the predicted action rates. We can also consider using combinations
of both CTR and those post-read action rates together to blend news
articles so that such a ranking can be potentially useful for users
not only clicking on the articles, but also sharing or commenting
them after reading.

However, to the best of our knowledge, little prior work provides a
thorough analysis of those post-read actions and our understanding
of them is very limited. For example, how indicative is the CTR of
an article to those post-read actions? Do users in different age
groups have different preferences for those post-read action types?
How difficult is it to predict the post-read action rates? To
answer these questions, we collect a data set from an online news
website and conduct an exploratory analysis upon that.
Interestingly, we found that those post-read actions are mostly
orthogonal to CTR. For example, the kinds of articles that users
like to share are quite different from those they like to read,
suggesting two sides of users' news consumptions: private and
public. Furthermore, our analysis also shows that different users
prefer different post-read actions but the signal-to-noise ratios
of those post-read actions are much lower than clicks. Thus
sophisticated models are required to model these post-read actions.




The main challenges in modeling these post-read actions are due to
data sparsity which is more severe than the sparsity of CTR
estimation because those post-read actions are conditional on
clicks. As far as we know,
little work in the area of news recommendation has considered the
use of such signals despite the availability of post-read behavior
data on most online news sites. In fact, an increasing number of
users are using social media to actively promote/demote news
articles through their circles and are freely expressing their
opinions through comments. Thus, our focus in this paper is on
machine learning techniques that can mitigate sparsity and provide
reliable estimates of post-read action rates that is essential to
incorporate such signals into news recommendation.

Fortunately, our analysis shows that positive correlations exist
among different types of post-read actions. Exploiting such
correlations helps in borrowing information across action types and
thus reduce the data sparsity. In this paper, we model correlations
among post-read action types through several variations of latent
factor models, including a novel variant for multivariate response,
and show that it is indeed feasible to estimate post-read action
rates reliably. We show that the novel variant proposed in this
paper outperforms other models in terms of a number of ranking
metrics. This opens the door to using multi-faceted ranking to
improve news recommendation on websites.

Our contributions are as follows. We conduct a thorough
exploratory analysis of the post-read actions and report several
interesting observations which support that multi-faceted ranking
of news articles is desirable. Furthermore, we propose a new
problem of estimating rates of different post-read action types
which provide a reasonable approach to multi-faceted news ranking
in the context of news recommendation.
In particular, we propose a novel Locally Augmented Tensor model
(LAT) that effectively explore the correlations in noisy and sparse
multivariate response data. We compare this model with a set of IR
models and several strong matrix factorization baselines and
experimentally show that the LAT model can significantly
improve news ranking accuracy in multiple facets.

\section{Exploratory Data Analysis} \label{sec:data}


\begin{figure}[t]\centering
\epsfig{file=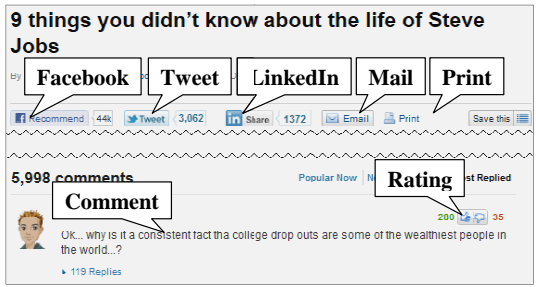,width=0.40\textwidth}
\caption{Illustration of post-read actions}
\label{fig:example}
\end{figure}

We study post-read behavior based on data collected from a major
news site in the US that obtains several million visits on a
monthly basis. Although this does not represent the entire news
reading population in the US and elsewhere in the world, it has a
large enough market share to study online news consumption behavior
in the US.
The site provides various functionalities for users to act after
reading an article. Figure~\ref{fig:example} shows a portion of a
typical news article page. On top are links/buttons that allow a
user to share articles on various social media websites such as
Facebook, Twitter and LinkedIn. The user can also share the article
with others or herself via email or by printing a hard copy. At the
bottom portion of the page, the user can leave comments on the
article or rate other users comments by thumb-up or thumb-down.

In addition to links/buttons that facilitate post-read actions,
most article pages on this site publish a module that recommends
interesting article links to users. This module is an important
source to create page views on the site and hence recommends
article links that maximize overall click-through rates. To
estimate the CTR of each article unbiasedly, a small portion of the
user visits are shown a random list of articles and the CTR
estimated from this small portion of traffic is used to perform our
CTR analysis.

\commentout{
The Yahoo! site also serves articles at random to a small fraction of randomly chosen user population on this module;
the click logs collected from this experimental population provides an unbiased source of click data that we shall exploit
in our analysis when necessary; we call this the \textit{random bucket}. Other sources that create page views on Yahoo! News are clicks on article links published in News category pages, Yahoo! front page, major search engines, and many others. News for You is our main source for studying click behavior in this paper, while post-read actions are based on all page views.
}

\subsection{Data}
We describe the news data analyzed in this paper.

\parahead{Source of Data}
We collect two kinds of data --- (1) all page views on the news
site to study post-read actions (these page views are generated via
clicks on links to news articles published by the site on the web);
and (2) click logs from the module as described earlier. To
distinguish link views on the module from page views of news
article pages (after clicking article links), we shall refer to the
former as ``linkview'' while the latter is referred to as
``pageview''.
For instance, using this terminology, pre-read article
click-through rate (CTR) is computed as the number of clicks
divided by the number of linkviews on the module and post-read
Facebook share rate (FSR) is computed as the number of sharing
actions divided by the number of pageviews.
Post-read action rates of other types can be computed similarly; we
focus on the following post-read actions: Facebook share, email,
print, comment, and rating.

\begin{figure}[t]
\centering
\epsfig{file=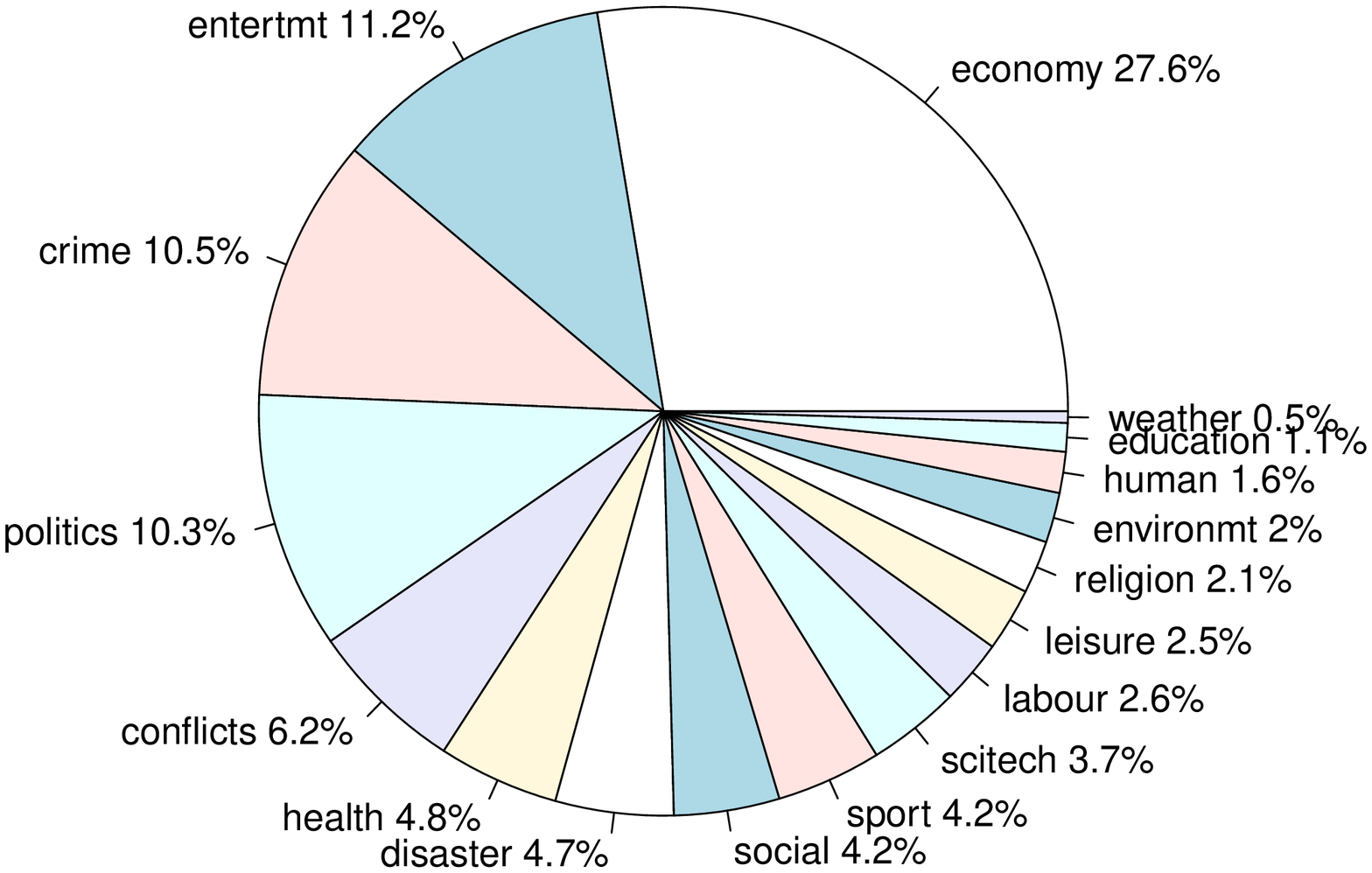,width=0.35\textwidth}
\\
\vspace{5mm}
\small
\begin{tabular}{|l|l|l|l|}

\hline Abbreviation & Full name\\
\hline economy & economy, business and finance \\
\hline entertmt & arts, culture and entertainment \\
\hline crime & crime, law and justice \\
\hline conflicts & unrest, conflicts and war \\
\hline disaster & disaster and accident\\
\hline social & social issue \\
\hline scitech & science and technology \\
\hline leisure & lifestyle and leisure \\
\hline religion & religion and belief \\
\hline environmt & environmental issue \\
\hline human & human interest\\
\hline
\end{tabular}
\caption{Distribution of news articles over categories}
\label{fig:itemcat}
\vspace{-0.15in}
\end{figure}

\parahead{Data Diversity}
The data used in our analysis was collected over a period of
several months in 2011. We selected articles that were shown on the
module and were clicked at least once, received at least one
comment and one post-read action type out of Facebook share, email
and print. This gives us approximately 8K articles that were
already classified into a hierarchical directory by the publishers.
We use the top three levels of the hierarchy for our analysis. The
first level of the hierarchy has 17 categories; the distribution of
article frequency in these categories is shown in
Figure~\ref{fig:itemcat}. As evident from this figure, news
articles published on the site are diverse in nature and provides a
good source to study user interaction with online news. We also
obtain user demographic information which includes age, gender and
geo-location (identified through IP address). All user IDs are
anonymized. In total, we have hundreds of millions of pageview
events in our data which are sufficient for us to estimate the
post-read action rates.



\subsection{Pre-Read vs Post-Read}
\commentout{
\begin{figure*}[t]\centering
\epsfig{file=./response-corr-by-item-overall.eps,width=0.8\textwidth} %
\vspace{-0.15in}
\caption{Scatter plots between click and other actions types (each dot represents an article). Only articles
with at least 2k linkviews and pageviews were considered.}
\label{fig:ctr-vs-others} %
\end{figure*}
}
In this section, we investigate the relationship between pre-read
(click) and post-read actions.
For example, is a highly clicked article also highly shared or
commented by users? For each article, we compute the article's
overall click-through rate (CTR) on the module and post-read action
rates of different types.
In Figure~\ref{fig:corr-actions}, we show the correlation between
clicks and other actions types using Pearson's correlation (the
first column or the last row).
We observe very low correlation between click rates and other
post-read action rates. We also computed the correlations after
stratifying articles by categories and found that the correlations
are still very low. This lack of correlation is perhaps not
surprising: clicks are driven by user's topical interest in certain
articles vs others, while post-click behaviors are inherently
conditioned on clicks and hence topical interest. Hence, ranking
articles using CTR and other post-click indicators would perhaps
lead to different rankings. For instance, if the goal of a news
website is to maximize CTR but also ensure a certain minimum number
of tweeting and it is possible to predict articles that are more
likely to be tweeted, the rankings could be altered based on CTR
and tweeting rates to achieve such an objective.

\commentout{
\parahead{How to measure article popularity?}
Article popularity measured in terms of article CTR is not consistent
with popularity in terms of other post-read action rates. Of course, some or all of the
inconsistencies may be due to differences in user populations
who click and those who engage with post-read actions.  In instances
of content recommendation where articles are ranked based on CTR popularity; this analysis provides more food for thought; perhaps
blending recommendations based on different popularity measures could lead to
better user experience!
}

\subsection{Correlations among Post-Read Actions}

\begin{figure}[t]\centering
\epsfig{file=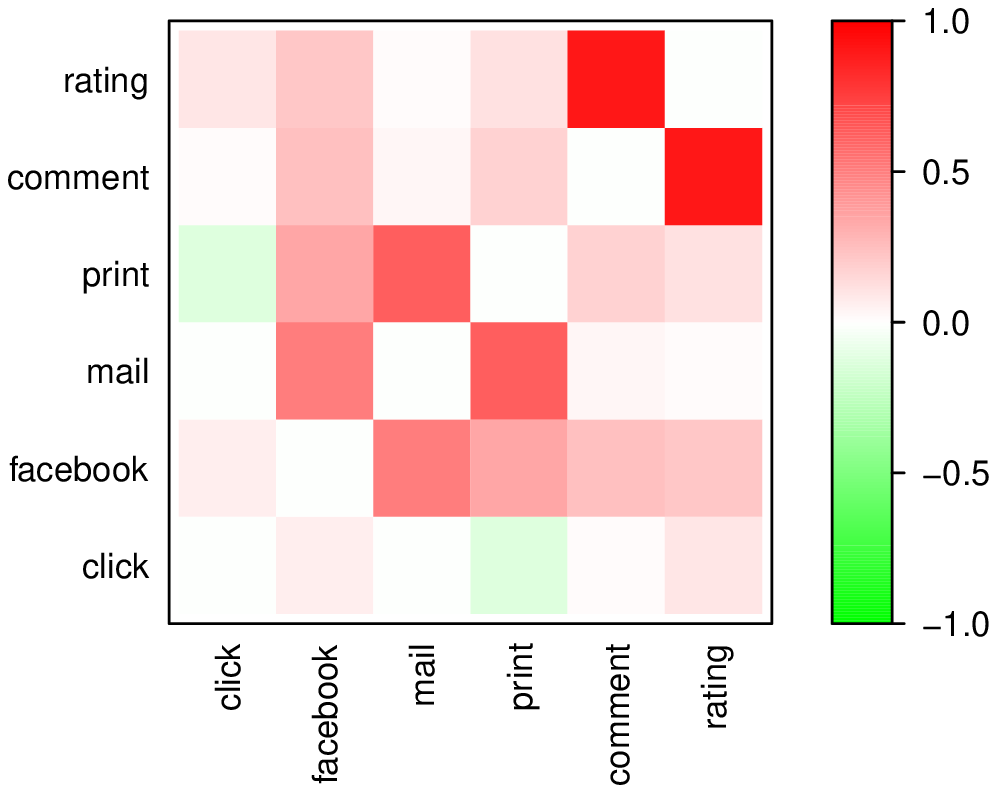,width=0.30\textwidth} %
\caption{Correlation between different action types (diagonals cells are not of interest)} %
\label{fig:corr-actions} %
\vspace{-0.2in}
\end{figure}

In Figure~\ref{fig:corr-actions}, we also show all pairwise
Pearson's correlations among post-read action types, computed using
article-level action rates. \commentout{ Notice that the diagonal
cells represent self correlation, which is always 1 and not of
interest;  thus, we simply give them no color. A green color means
that the two are negatively correlated and a red color means that
the two are positively correlated. In this figure, we omit the self
correlation. From this figure, we can see again that click has very
low correlation with post-read actions. } We observe positive
correlations among various post-read action types; Mail has high
correlation with Facebook and print, but not with comment and
rating. There is high correlations among Facebook, mail, and print.
Not surprisingly, comment and rating are also highly correlated.
These provides evidence of being able to leverage correlations
among post-read action types to improve estimation.

A word of caution: it is not necessary that correlations will hold
when the data is disaggregated at the (user, item) level since our
data is observational and subject to various sources of bias. It is
not possible to study correlations at the (user, item) level
through exploratory analysis due to lack of replicates; we will
study this problem rigorously through a modeling approach described
in Section~\ref{sec:model}. The exploratory analysis is shown to
provide a flavor of our data (since we are not able to release it
due to reasons of confidentiality) and to gain some insights at an
aggregate level.

\subsection{Read vs. Post-Read: Private vs. Public}

We now compare users' reading behavior with their post-read
behavior. Specifically, is post-read behavior uniform across
different article types or user types? Does Joe, a typical young
male from California comment and share most of the articles he
reads?

\begin{figure*}[t]\centering
\centering \vspace{-0.25in} \hspace{-0.4in} %
\subfigure[On top level categories]{\label{fig:post-pageview}
	\raisebox{5mm}{\epsfig{file=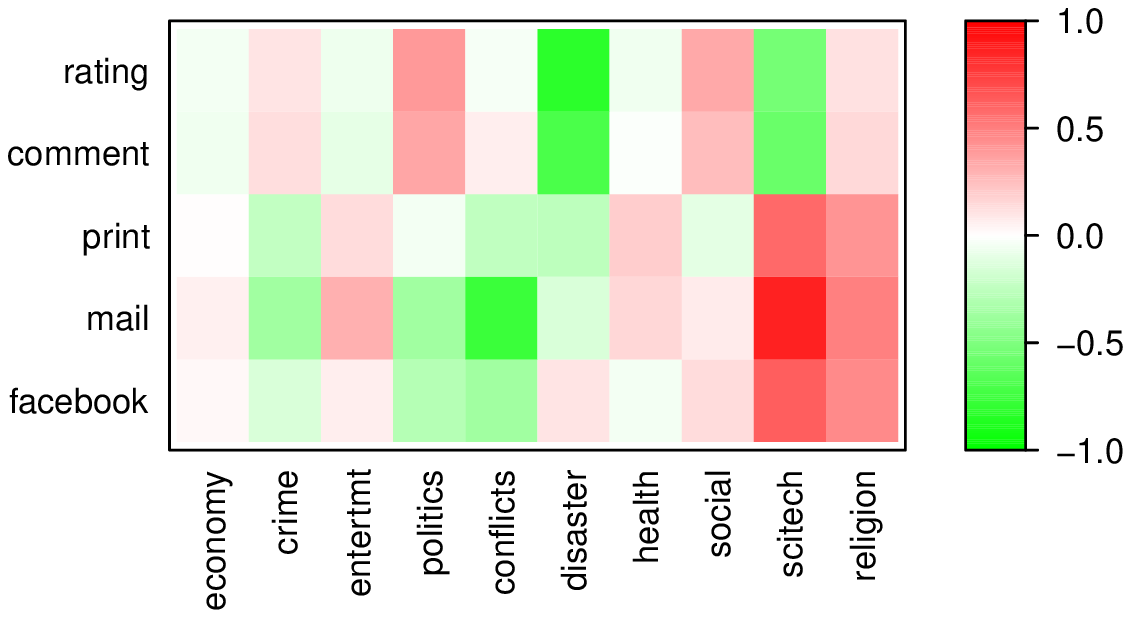,width=0.4\textwidth}} %
}%
\subfigure[On subcategories of
economy]{\label{fig:post-economy-pageview}
    \epsfig{file=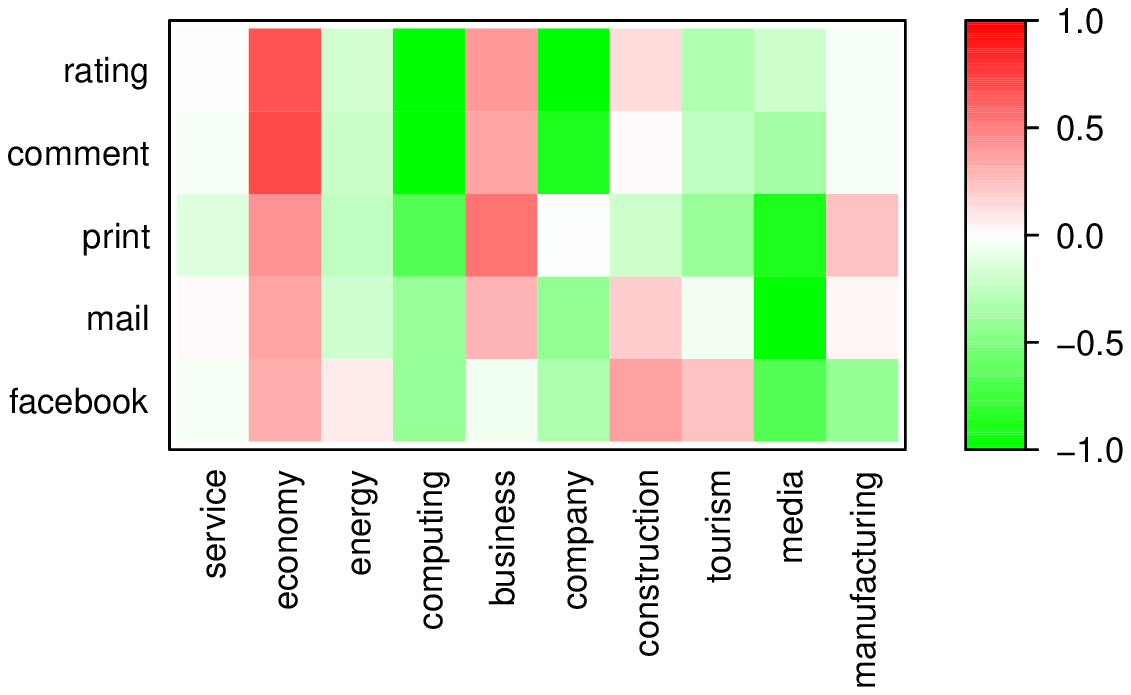,width=0.4\textwidth} %
}%
\caption{Difference between pageview and post-read actions. This also shows
variation of post-read action rates of different types by article categories} %
\label{fig:post-variation}
\vspace{-0.05in}
\end{figure*}

To understand this, we use a vector of the fractions of pageviews
in different article categories to represent the reading behavior.
One can think of this vector as a multinomial probability
distribution over categories; i.e., the probability that a random
pageview is in a given category.  Similarly, marginal post-read
behavior of an action type in a category is represented as a vector
of fractions of post-read actions of that type in that category. To
compare a post-read behavior vector with the reading behavior
vector, we compute the element-wise ratio between the two vectors.
Figure~\ref{fig:post-pageview} shows these ratios on the log-scale
using the top 10 most viewed categories, where the categories are
ordered according to the numbers of pageviews they received
(highest on the left).  All the sample sizes are sufficiently large
(with at least tens of thousands of post-read actions) to ensure
statistical significance. To help understand this plot, let us
consider the green color (i.e., negative value) in the (mail,
conflicts) cell for instance. It indicates that a typical user is
more likely to read an article about conflicts than email it.
In general, if post-read action behavior of users were the same as
reading behavior or uniform across news types, the ratios (on
log-scale) should be clustered around $0$.
Obviously this is not the case for all action types as we see both
``hot'' and ``cold'' cells in the plot.

Some interesting tidbits. Users are more likely to read articles
about crime, politics and conflicts than to share them with friends
via email or on Facebook; they are more likely to read about
disaster and science \& technology but reluctant to comment on
them. When it comes to science and religion, they are eager to
share more.  They are also more open to leave comments and engage
in discussions in a public forum on matters of politics.

\commentout{
 We make a few observations:
\myItemizeBegin
\item Sharing-related actions (facebook and mail) have similar behavior. It is interesting to see that users like to read articles about crime, politics and conflicts (as these categories are among the top 5 most viewed categories), but not so much to share them.
\item Opinion-related actions (comment and rating) have similar behavior.  It is interesting to see that users read articles about disaster and science \& technology, but tend not to express opinions on them.
\item Science \& technology, politics and crime are the main categories that show significant differences between sharing and opinion-expressing.
\myItemizeEnd
}
We observe an interesting pattern in news consumption. Reading news
articles is a private activity, while sharing (Facebook and mail)
or expressing opinions (comment and rating) on articles is a public
activity and there is difference in a typical user's public and
private activity. Users tend to share articles that earn them
social prestige and credit but they do not mind clicking and
reading some salacious news occasionally in private.

\subsection{Variation in Post-Read Action Rates at Different Resolutions}
\label{sec:var-post-read} The previous subsection showed
interesting differences in read and post-read behavior across
different article types. In this section, we study variation in
post-read action rates by slicing and dicing the data at different
resolutions. We note that analysis at some coarse resolution for
data obtained through a non-randomized design may not reveal the
entire picture; ideally such inferences should be drawn after
adjusting for data heterogeneity at the finest, i.e, (user, item)
resolution. It is impossible to study variation at this fine
resolution through exploratory analysis due to lack of replicates.
Our goal in this section is to study variation at resolutions where
enough replicates exist. Such an analysis also provide insights
into the hardness of predicting action rates and whether
sophisticated modeling at fine resolutions is even necessary. For
instance, if all science articles behave similarly, it is not
necessary to model data at the article level within the science
category.


\parahead{Variation across article categories} To study variation in
post-read action rates across article categories, we compute the
ratio between the category-specific post-read action rates (i.e.,
\#actions in the category divided by \#pageviews in the category)
and the global action rate (i.e., total \#actions divided by total
\#pageviews) using the top 10 most viewed categories for each
action type. This is exactly what Figure~\ref{fig:post-pageview}
and Figure~\ref{fig:post-economy-pageview} show. As we noted
earlier, there is variation in action rates at this resolution as
evident from the ``hot'' and ``cold'' cells. \commentout{ It is
interesting to observe that users have a high propensity to share
articles about science \& technology and religion, and to comment
and rate comments on articles about politics and social issues. The
economy category shows neutral patterns but it is a big category.
After zooming further at the second level of the category
hierarchy, we also see variations in
Figure~\ref{fig:post-economy-pageview}. For instance, there is
positive propensity for general economy articles for all action
types, especially for comment and ratings. We only selected
categories with large sample sizes in these plots to avoid showing
patterns based on small data.  These heat maps clearly show that
article categories are able to explain variations in post-read
action rates. }

\begin{figure}[t]\centering
\epsfig{file=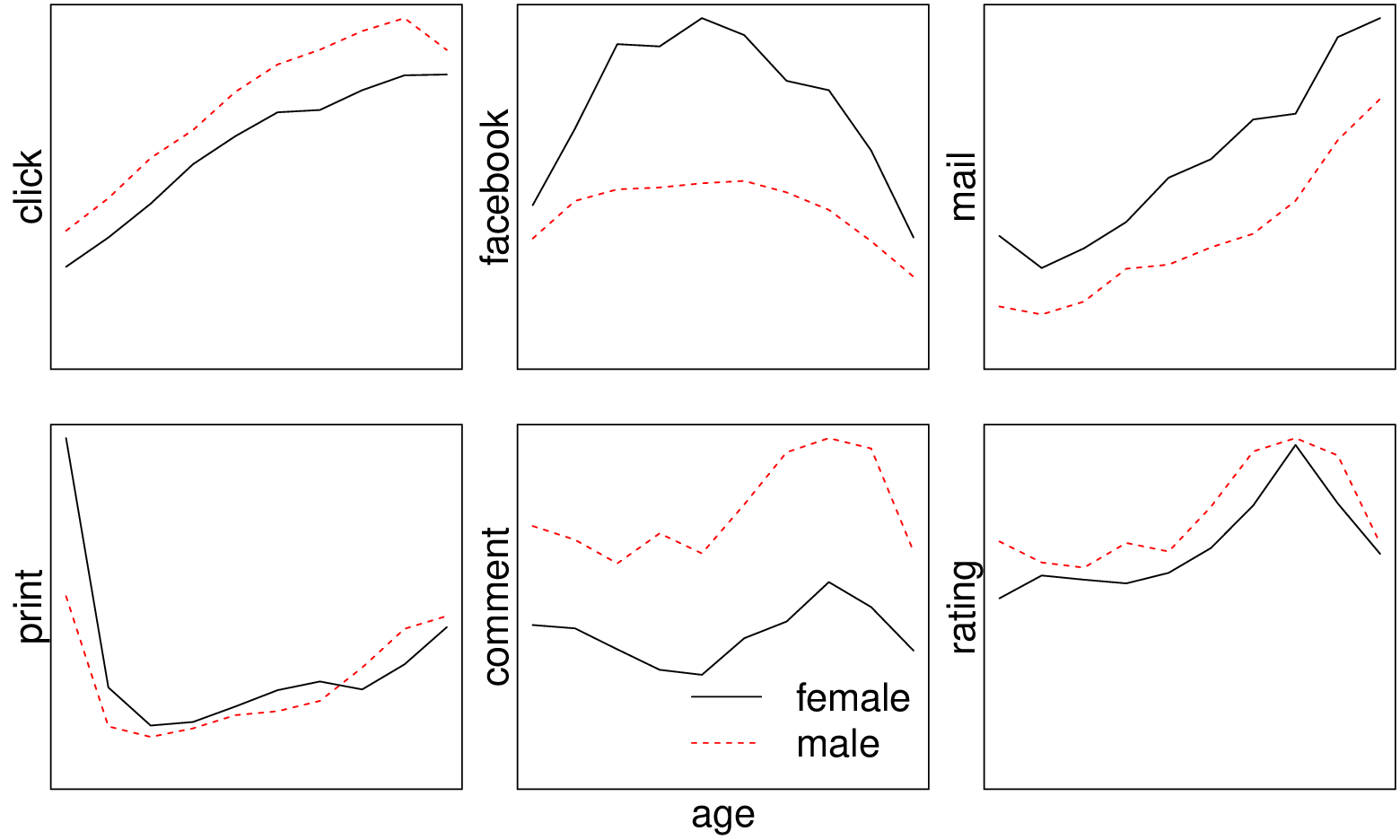,width=0.45\textwidth} %
\caption{Post-read action rate variation over age-gender segments} %
\label{fig:age-gender} %
\end{figure}

\parahead{Variation across user segments}
We segment users by age and gender and show post-read action rates
for the two genders across different age-groups in
Figure~\ref{fig:age-gender}. Once again we see variation. Some
interesting observations:  Facebook share rates are highest among
young and middle aged users. Users in older age groups tend to mail
more but young users tend to share more on Facebook, and also print
more. We see females to have surprisingly high share rates and male
users tend to comment more on articles. We also include pre-read
click actions in this figure and observe that users in older age
groups tend to click more; males across all age-groups are more
active clickers than females.

\parahead{Variation within categories and segments}
We now dig deeper and analyze variation at the article resolution after stratifying our data by article categories and user segments.
High within-category/segment variations at the article level
indicate excessive heterogeneity with categories and segments and
suggest the need to model the rates at finer resolutions. To study
such variation, we leverage the coefficient of variation defined as
$\sigma/\mu$, where $\sigma$ is the standard deviation of article
action rates within a given category (or category $\times$ user
segment) and $\mu$ is the average article action rate in the
category (or category $\times$ user segment). $\sigma/\mu$ is a
positive number; smaller values indicate less variation. In
general, values above $0.2$ are indicative of high variation.

\begin{figure}[b]\centering
\centering \vspace{-0.25in} \hspace{-0.3in} %
\subfigure[On article categories]{\label{fig:sd-m-cat}
	\epsfig{file=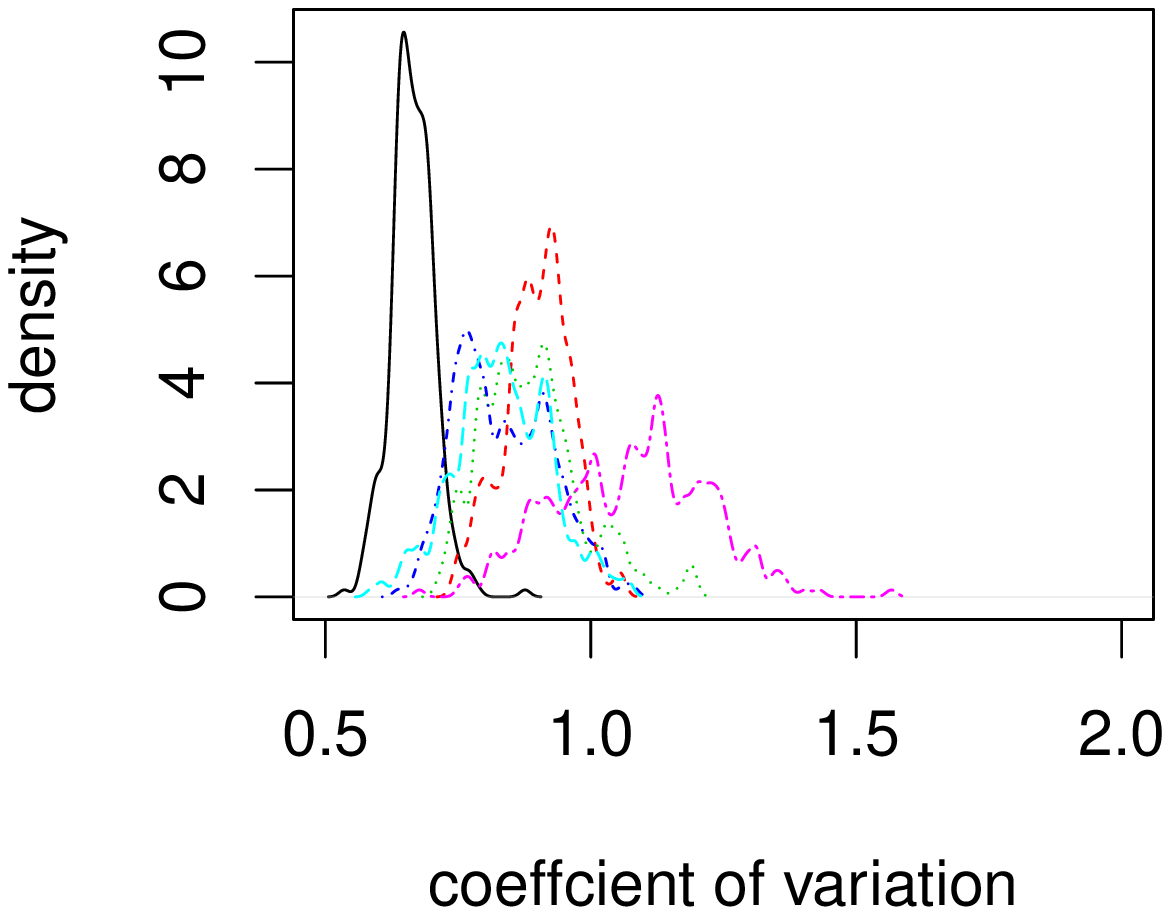,width=0.255\textwidth} %
}%
\subfigure[On categories $\times$ age-gender]{\label{fig:sd-m-agender-cat}
    \epsfig{file=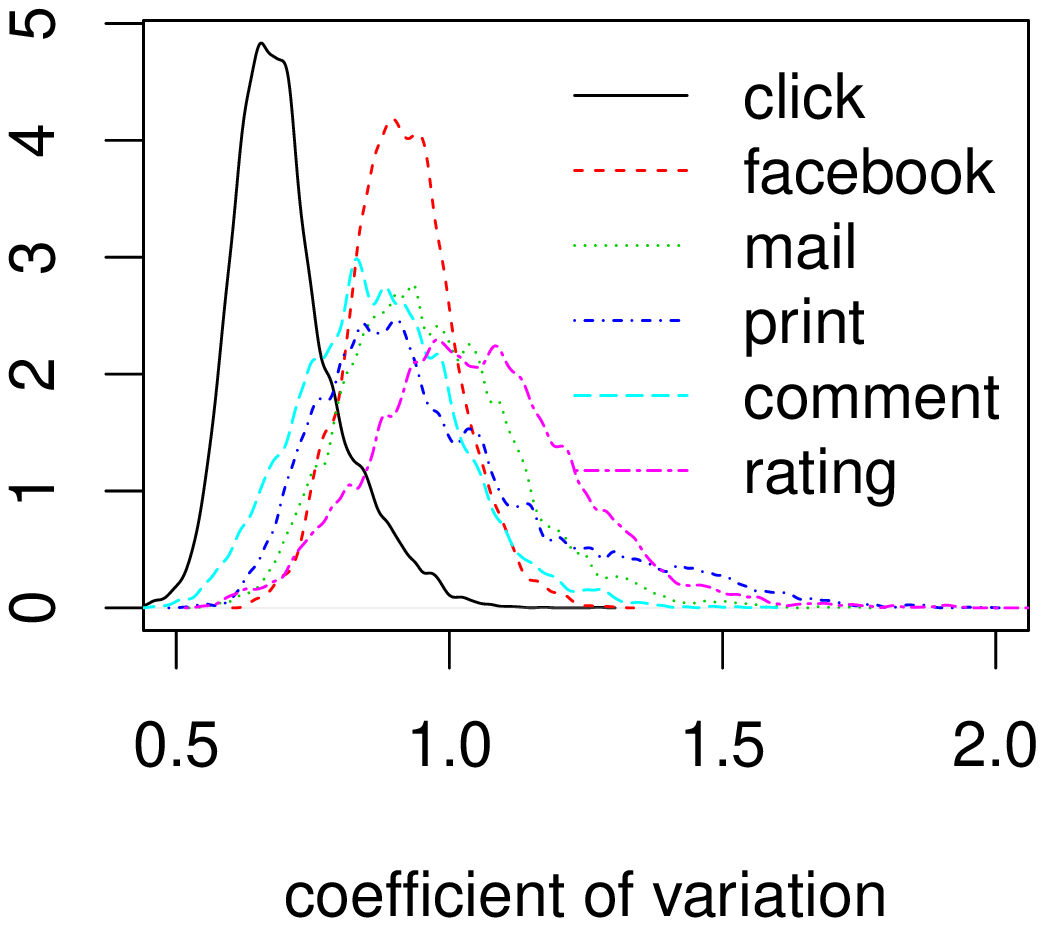,width=0.23\textwidth} %
}%
\caption{Density of coefficient of variation} %
\label{fig:density-coefficent} \vspace{-0.05in}
\end{figure}

In Figure~\ref{fig:sd-m-cat} and
Figure~\ref{fig:sd-m-agender-cat}, we show the distribution of
coefficient of variation with respect to article categories and the
cross-product of categories with user age-gender. From these two
figures, we can see that all post-read actions have much larger
coefficients of variation than click. This means that although
there is variation in average post-read behavior across categories
and user segments, the variation at the article resolution within
each such stratum is high making it difficult to predict article
post-read action rates than article click rates based on the
category information. Comparing the two figures, we can see that
adding user features helps little in terms of reducing coefficients
of variation indicating that stratification by user segments does
not help in explaining article level variation within each
category. Perhaps users with a given (age, gender) segments have
different news consumption behavior at the article level.

Our exploratory analysis suggests that predicting post-read actions
of any type is harder than estimating CTR. We also see that while
using features like article category and user demographics are
useful, there is heterogeneity at the article and user resolution
that has to be modeled. We also see evidence of positive
correlations among post-read action types; it is interesting to
study if such correlations can make the estimation task any better.
We explore such an approach by modeling all post-read action types
simultaneously at the finest (user, item) resolution in
section~\ref{sec:model}.

\section{Post-Read Action Prediction}
\label{sec:model}

In this section, we present our locally augmented tensor (LAT) model
for predicting users' post-read actions.  Given a user, an item (i.e.,
article) and a post-read action type (e.g., commenting, Facebook
sharing), we want to predict whether the user would take the action
after reading the item. The main challenges are:
\myItemizeBegin

\item {\it Data sparsity:} Post-read actions are rare events. Most
  users only have single-digit post-read actions in a month in our dataset.
  If we further breakdown actions by types, data becomes sparser.

\item {\it Diverse behavior across action types:} As we saw in
  Section~\ref{sec:data}, users behave differently for different
  action types.  For example, the kinds of articles that users like to
  share are quite different from those that they like to comment on.
\myItemizeEnd

To handle data
sparsity, an attractive approach is to appropriately pool the action
data of a user from all types, so that the action data of one type can
be used to improve the prediction performance for another type.
However, naive ways of pooling action data that ignore the differences
between action types may lead to poor performance, especially for
our sparse post-read data.

\parahead{Problem definition} Consider an online news system with $M$
users, $N$ items and $K$ post-read action types.  Let $y_{ijk}$ denote
whether user $i$ takes a post-read action of type $k$ on item $j$. If
the user takes the action, $y_{ijk} = 1$; if the user reads the item
and does not take the action, $y_{ijk} = 0$; if the user does not read
the item, $y_{ijk}$ is unobserved.  We also call $y_{ijk}$ the
observation or response of user $i$ to item $j$ of type $k$.  Each
user is associated with a feature vector (e.g., age, gender,
geo-location).  Each item is also associated with a feature vector
(e.g., content categories, words and entities in the item).  Because
$i$ always denotes a user and $j$ always denotes an item, we slightly
abuse our notations by using $\bm{x}_i$ to denote the feature vector of
user $i$ and $\bm{x}_j$ to denote the feature vector of item $j$.
Given user features, item features and a set of training observations,
our goal is to predict the response of a set of (user, item, action
type) triples that do not appear in the training data.

We model the data using variants of factor models. We begin with a review of baseline matrix factorization models and then extend them to address the above challenges.

\subsection{Baseline Factor Models}
\label{sec:baseline-model}

Matrix factorization is a popular method for predicting user-item
interaction.
User-item
interactions can be represented through a $M \times N$ matrix $\mathbf{Y}$, where the value
$y_{ij}$ of the $(i,j)^{th}$ entry is the response of user $i$ to item $j$.
Notice that this is a matrix with many unobserved (i.e., missing)
entries because a user typically does not interact with many items.
The main idea of matrix factorization is to
obtain two low rank matrices $\mathbf{U}_{M \times F}$ and $\mathbf{V}_{N \times F}$ such
that $\mathbf{Y}$ is close to the product $\mathbf{U} \mathbf{V}'$ measured in terms of a loss
function $l(\mathbf{Y},\mathbf{U} \mathbf{V}')$ (e.g. squared-error, logistic). Here $F$ is
much smaller than $M$ and $N$. Such a decomposition enables us to predict the unobserved entries
in the response matrix.

Each row $\bm{u}_i$ of matrix $\mathbf{U}$ is called the factor vector
of user $i$, representing his/her latent profile.  Similarly, each row
$\bm{v}_j$ of matrix $\mathbf{V}$ represents the latent profile of item $j$.
Intuitively, the inner product $\bm{u}'_i \bm{v}_j$ is
a measure of similarity between the profiles of user $i$ and item $j$,
representing how much $i$ likes $j$.  It is common to also add a bias
term $\alpha_i$ for each user $i$ to represent his/her average
response to items, and a bias term $\beta_j$ for each item $j$ to represent its
popularity.  Then, the response $y_{ij}$ of user $i$ to item $j$ is
predicted by $\hat{y}_{ij}=\alpha_i + \beta_j + \bm{u}'_i \bm{v}_j$.

Let $\ell\ell(y,x) = -\log(1+\exp\{-(2y-1)\,x\})$ denote the logistic log-likelihood for a binary observation $y$. The loss function is given by
\begin{equation}
\label{eqn:lgtloss}
l(\mathbf{Y},\mathbf{U} \mathbf{V}') = -\sum_{\textrm{observed~} (i,j)} \ell\ell(y_{ij},\hat{y}_{ij})
\end{equation}
Optimizing the loss function in Equation~\ref{eqn:lgtloss} tends to give estimates that overfit sparse data since the number of parameters is too large even for small values of $F$.
It is customary to impose penalty (regularization) to avoid overfitting, the most commonly
used penalty is to constrain the $L_2$ norm of parameters. Thus, we obtain parameter estimates
by minimizing
\begin{equation}
\label{eqn:sgd}
\textstyle
l(\mathbf{Y},\mathbf{U} \mathbf{V}')+\frac{1}{2\sigma^2_{\alpha}}\sum_{i}\alpha^{2}_{i} + \frac{1}{2\sigma^2_{\beta}}\sum_{j}\beta^{2}_{j}+ \frac{1}{2\sigma^2_{u}}\sum_{i}||u_{i}||^{2}+ \frac{1}{2\sigma^2_{v}}\sum_{j}||v_{j}||^{2}
\end{equation}
where the $\sigma^2_\cdot$s are tuning constants. Stochastic gradient descent (SGD) is a popular method to perform such optimization. However, our models involve several tuning constants that are hard to estimate
using procedures like cross-validation.  Further, SGD also requires tuning learning rate parameters.
Thus, we pursue a different estimation strategy
for fitting our models by working in a probabilistic framework and using a Monte-Carlo Expectation Maximization (MCEM) procedure. The MCEM approach we follow is both scalable and {\em estimates all parameters
automatically} through the training data.

\subsection{Probabilistic Modeling Framework}

\parahead{Observation model}
Matrix factorization can also be interpreted in a probabilistic modeling framework.  The given $y_{ij}$s are the observations, based on which we want to estimate the unobserved latent factors $\alpha_i$, $\beta_j$, $\bm{u}_i$ and $\bm{v}_j$.  For numerical response, it is common to use a Gaussian model.
\begin{equation*}
y_{ij} \sim N(\alpha_i + \beta_j + \bm{u}'_i \bm{v}_j,~ \sigma^2_{y}),
\end{equation*}
where $N(\mu, \sigma^2)$ denote a Gaussian distribution with mean $\mu$ and standard deviation $\sigma^2$.  For binary response, it is common to use a logistic model.
\begin{equation*}
y_{ij} \sim \textit{Bernoulli}(p_{ij}) \textrm{  and  }
\textstyle \log\frac{p_{ij}}{1 - p_{ij}} = \alpha_i + \beta_j + \bm{u}'_i \bm{v}_j.
\end{equation*}
For ease of exposition, we use $y_{ij} \sim \alpha_i + \beta_j + \bm{u}'_i \bm{v}_j$ to denote that $y_{ij}$ is predicted based on $\alpha_i + \beta_j + \bm{u}'_i \bm{v}_j$
either using the Gaussian model or the logistic model.

\parahead{Regression priors}
Although $\bm{u}_i$ and $\bm{v}_j$ are low dimensional, there are
still a large number of factors to be estimated from sparse data,
which can similarly lead to overfitting.  A common approach is to
shrink the factors toward zero; i.e., if a factor is not supported
by enough data, it's estimated value should be close to zero.  When
features are available, we can achieve better performance by
shrinking the factors toward values predicted by features, instead
of zero.  For example, if user $i$ has very few activities in the
training data, instead of ensuring $\bm{u}_i$ to be close to zero,
we predict $\bm{u}_i$ using a regression function $G \bm{x}_i$,
where $G$ is the regression coefficient matrix learned from
training data through linear regression.  Notice that $\bm{u}_i$ is
a vector; thus, $G$ is a matrix, instead of a vector.  If features
are predictive, then we can obtain good $\bm{u}_i$ estimates even
for users without any training data.  Specifically, we assume the
following priors.
\begin{equation*}
\begin{array}{rl rl}
\alpha_i & \sim N(\bm{g}' \bm{x}_i, \sigma^2_\alpha), & ~~~~~~~
\bm{u}_i & \sim N(G \bm{x}_i, \sigma^2_u), \\
\beta_i  & \sim N(\bm{d}' \bm{x}_j, \sigma^2_\beta), & ~~~~~~~
\bm{v}_j & \sim N(D \bm{x}_j, \sigma^2_v).
\end{array}
\end{equation*}

\parahead{Training and prediction}
We defer the training algorithm to Section~\ref{sec:training-algo}.  Here, we note that this model is a generative model that specifies how the observations $y_{ij}$ are generated according to the latent factors $\alpha_i$, $\beta_j$, $\bm{u}_i$ and $\bm{v}_j$, which in turn are generated according to the prior parameters ($\bm{g}$, $G$, $\bm{d}$, $D$ and the $\sigma^2_\cdot$s).  Given a set of observations, we first obtain the maximum likelihood estimate (MLE) of the prior parameters.  Then, based on the MLE of prior parameters and the observations, we obtain the posterior mean of $\alpha_i$, $\beta_j$, $\bm{u}_i$ and $\bm{v}_j$, which then can be used to predict the response of an unseen $(i,j)$ pair by $\alpha_i + \beta_j + \bm{u}'_i \bm{v}_j$.


\parahead{Baseline models}
Two straightforward ways of applying matrix factorization to our problem are as follows:
\begin{itemize}
\item {\it Separate Matrix Factorization (SMF):}  Treat observations of $K$ action types as $K$ separate matrices and apply factorization to each of them independently; i.e.,
$$
y_{ijk} \sim \alpha_{ik} + \beta_{jk} + \bm{u}'_{ik} \bm{v}_{jk}.
$$
\item {\it Collapsed Matrix Factorization (CMF):}  Collapse observations of all types into a single matrix and apply factorization to it; i.e.,
$$
y_{ijk} \sim \alpha_{i} + \beta_{j} + \bm{u}'_{i} \bm{v}_{j},
$$
where the right hand side does not depend on type $k$.
\end{itemize}
Notice that SMF is a strong baseline because for users and items with
large number of training samples, their factors can be estimated accurately. For
users and items without much training data, their factors can still be
predicted by features. Compared to CMF, SMF has $K$ times more factors
to be estimated from data and is more sensitive to data sparsity.
Although CMF is less sensitive to data sparsity, it ignores the behavioral differences across different
action types, which may lead to bias and poor performance.

\subsection{Locally Augmented Tensor Model}
\label{sec:lat-model}

We now introduce the locally augmented tensor (LAT) model, which
addresses data sparsity through tensor factorization, augmented with
SMF to model the residuals locally for each action types.  We first
specify the model and then discuss how it works.

\parahead{Model specification}
The action $y_{ijk}$ that user $i$ takes on item $j$ of type $k$ is modeled as:
\begin{align}
y_{ijk} & \sim \alpha_{ik} + \beta_{jk} +
	\left<\bm{u}_i, \bm{v}_j, \bm{w}_k\right> + \bm{u}'_{ik} \bm{v}_{jk},
\label{eq:lat}
\end{align}
where $\left<\bm{u}_i, \bm{v}_j, \bm{w}_k\right> = \sum_{\ell} \bm{u}_i[\ell]\, \bm{v}_j[\ell]\, \bm{w}_k[\ell]$ is the tensor product of three vectors $\bm{u}_i$, $\bm{v}_j$ and $\bm{w}_k$, and $\bm{u}_i[\ell]$ denotes the $\ell$th element in vector $\bm{u}_i$.  The intuitive meaning of the factors are as follows.
\myItemizeBegin
\item $\alpha_{ik}$ is the type-specific bias of user $i$.
\item $\beta_{jk}$ is the type-specific popularity of item $j$.
\item $\left<\bm{u}_i, \bm{v}_j, \bm{w}_k\right>$ measures the similarity between user $i$'s global profile $\bm{u}_i$ and item $j$'s global profile $\bm{v}_j$ weighted by a type-specific weight vector $\bm{w}_k$.  These profiles are called global because they are not type-specific.  Note that this weighted inner product (i.e., tensor product) imposes a constraint when we try to use it to approximate the observations $y_{ijk}$.  Specifically, it may not be flexible enough to accurately model post-read actions when there is diverse behavior across action types.  However, this constraint in the parametrization helps to avoid overfitting when data is sparse.
\item $\bm{u}'_{ik} \bm{v}_{jk}$ also measures the similarity between user $i$ and item $j$ for type $k$ and is more flexible than the tensor product.  Thus, the residuals that the tensor product does not capture can be captured by this inner product of type-specific user factor $\bm{u}'_{ik}$ and item factor $\bm{v}_{jk}$.
\myItemizeEnd
To contrast the global factors $\bm{u}_i, \bm{v}_j$, we call the type-specific factors $\bm{u}_{ik}, \bm{v}_{jk}$ {\em local factors}.  Since we augment the tensor product with the inner product of local factors, the resulting model is called the locally augmented tensor model.
The priors of the factors are specified as follows.
\begin{align}
\alpha_{ik} & \sim N(\bm{g}_{k}^\prime \bm{x}_{ik} + q_{k} \alpha_i, ~\sigma_{\alpha,k}^2),
	~~~~ \alpha_i \sim N(0, 1) \label{eq:alpha} \\
\beta_{jk} & \sim N(\bm{d}_{k}^\prime \bm{x}_{jk} + r_{k} \beta_j, ~\sigma_{\beta,k}^2),
	~~~~ \beta_j  \sim N(0, 1) \label{eq:beta} \\
\bm{u}_{ik} & \sim N(G_k \bm{x}_i, \,\sigma_{uk}^2 I), ~~~~
\bm{v}_{jk} \sim N(D_k \bm{x}_j, \,\sigma_{vk}^2 I), \label{eq:local-uv} \\
\bm{u}_{i} & \sim N(\bm{0}, \,\sigma_{u0}^2 I),
	~ \bm{v}_{j} \sim N(\bm{0}, \,\sigma_{v0}^2 I),
	~ \bm{w}_{k} \sim N(\bm{0}, I), \label{eq:uvw}
\end{align}
where $\bm{g}_k$, $q_k$, $\bm{d}_k$, $r_k$, $G_k$ and $D_k$ are regression coefficient vectors and matrices similar to those discussed in Section~\ref{sec:baseline-model}.  These regression coefficients are to be learned from data and provide the ability to make predictions for users or items that do not appear in training data.  The factors of these new users or items will be predicted based on their features through regression.

\parahead{Training and prediction}
Given training data $\bm{y} = \{y_{ijk}\}$, the goal of the training process is to learn the latent factors $\bm{\eta} = \{\alpha_{ik}, \beta_{jk}, \alpha_{i}, \beta_{j}, \bm{u}_{i}, \bm{v}_{j}, \bm{w}_{k}, \bm{u}_{ik}, \bm{v}_{jk}\}$ and prior parameters
$\bm{\Theta} = \{\bm{g}_k, \bm{d}_k, q_k, r_k, G_k, D_k,$ the $\sigma_{\cdot}^2$s$\}$ (which consists of regression coefficients and variances) from the data $\bm{y}$.
The training algorithm will be given later.
After training, given an unobserved (user $i$, item $j$, action type $k$) triple, we predict the response as follows.  If both user $i$ and item $j$ have some type-$k$ observations in the training data, we just use their learned factors to make a prediction as $\alpha_{ik} + \beta_{jk} + \left<\bm{u}_i, \bm{v}_j, \bm{w}_k\right> + \bm{u}'_{ik} \bm{v}_{jk}$.  If user $i$ appears in the training data but has no type-$k$ observation ($\alpha_i$ and $\bm{u}_i$ are available from training but not $\alpha_{ik}$ and $\bm{u}_{ik}$), then we first predict $\alpha_{ik}$ as $\bm{g}_{k}^\prime \bm{x}_{ik} + q_{k} \alpha_i$ and $\bm{u}_{ik}$ as $G_k \bm{x}_i$, and then use Equation~\ref{eq:lat} to predict the response $y_{ijk}$.  Other cases can be handled in a similar manner.

\parahead{Special cases -- SMF and BST}
If we set $\alpha_i$, $\beta_j$, $\bm{u}_i$, $\bm{v}_j$ and $\bm{w}_k$ to zero, we obtain the SMF model (defined in Section~\ref{sec:baseline-model}).  If we set $\bm{u}_{ik}$ and $\bm{v}_{jk}$ to zero, we obtain the bias-smoothed tensor (BST) model proposed in \cite{reputation:kdd11} for a multi-context comment-rating prediction problem; i.e.,
$$
y_{ijk} \sim \alpha_{ik} + \beta_{jk} + \left<\bm{u}_i, \bm{v}_j, \bm{w}_k\right>.
$$
Notice that LAT has many latent factors and parameters to be learned.  It may be
  sensitive to overfitting.  However, because of the regularization
  provided by the priors (Equation~\ref{eq:alpha} to \ref{eq:uvw}),
  overfitting can be prevented when the prior variances are
  appropriately learned.

\subsection{Training Algorithm}
\label{sec:training-algo}

Since SMF, CMF and BST are special cases of LAT, we only discuss the training algorithm for LAT.  Based on Equations~\ref{eq:lat} to \ref{eq:uvw}, the joint log-likelihood of $\bm{y}$ and $\bm{\eta}$ given $\bm{\Theta}$ is
{\small\begin{equation*}
\begin{split}
& \textstyle \log \Pr(\mathbf{y}, \bm{\eta} \,|\, \bm{\Theta}) = \textrm{ some constant } + \sum_{ijk} \ell\ell(y_{ijk}, \hat{y}_{ijk}) \\
&	\textstyle
	- \frac{1}{2} \sum_{ik}
				\big( \log \sigma_{\alpha,k}^2 +
				 (\alpha_{ik} - \bm{g}_k^\prime \bm{x}_{ik} - q_{k} \alpha_i)^2
				  / \sigma_{\alpha,k}^2 \big) - \frac{1}{2}\sum_{i} \alpha_i^2\\
&	\textstyle
	- \frac{1}{2}\sum_{jk}
				\big(\log \sigma_{\beta,k}^2 +
				(\beta_{jk} - \bm{d}_k^\prime \bm{x}_{jk} - r_{k} \beta_j)^2
				  / \sigma_{\beta,k}^2\big) - \frac{1}{2}\sum_{j} \beta_j^2\\
&	\textstyle
	- \frac{1}{2}\sum_{ik}
				\big( F\log\sigma_{u,k}^2 +
				 \|\bm{u}_{ik} - G_k\bm{x}_i\|^2 / \sigma_{u,k}^2 \big)
	- \frac{1}{2}\sum_{k} \| \bm{w}_k \|^2\\
&	\textstyle
	- \frac{1}{2}\sum_{jk}
				\big( F\log\sigma_{v,k}^2 +
				 \|\bm{v}_{jk} - D_k\bm{x}_j\|^2 / \sigma_{v,k}^2 \big)\\
&	\textstyle
	- \frac{1}{2}\sum_{i} (F\,\log \sigma_{u}^2 + \|\bm{u}_i\|^2/\sigma_{u0}^2)
   - \frac{1}{2}\sum_{j} (F\,\log \sigma_{v}^2 + \|\bm{v}_j\|^2/\sigma_{v0}^2),
\end{split}
\end{equation*}}
where $\hat{y}_{ijk}=\alpha_{ik} + \beta_{jk} + \left<\bm{u}_i, \bm{v}_j, \bm{w}_k\right> + \bm{u}'_i\bm{v}_j$.
The goal of training is to obtain MLE of $\bm{\Theta}$; i.e.,
$$
\arg\max_{\bm{\Theta}} \Pr(\bm{y} \,|\, \bm{\Theta})
= \arg\max_{\bm{\Theta}} \int \Pr(\bm{y}, \bm{\eta}\,|\, \bm{\Theta}) \,d \bm{\eta},
$$
which can be obtained using the MCEM algorithm~\cite{booth99}.  The MCEM algorithm iterates between an E-step and an M-step until convergence.  Let $\hat{\bm{\Theta}}^{(t)}$ denote the current estimated value of the set of prior parameters $\bm{\Theta}$ at the beginning of the $t$th iteration.
\myItemizeBegin
\item {\bf E-step:}  We take expectation of the complete data log likelihood with respect to the posterior of latent factors $\bm{\eta}$ conditional on the
observed training data $\bm{y}$ and the current estimate of $\bm{\Theta}$; i.e., compute
$$
f_t(\bm{\Theta}) = E_{\bm{\eta} \sim \Pr\left(\bm{\eta} \,|\, \bm{y},\hat{\bm{\Theta}}^{(t)}\right)} [\log \Pr(\bm{y}, \bm{\eta} \,|\, \bm{\Theta})]
$$
as a function of $\bm{\Theta}$, where the expectation is taken over the posterior distribution of $(\bm{\eta} \,|\, \bm{y}, \hat{\bm{\Theta}}^{(t)})$, which is not in closed-form, thus, approximated by Monte Carlo mean through Gibbs sampling.

\item {\bf M-step:} We maximize the expected complete data log likelihood from the E-step to
obtain updated values of $\bm{\Theta}$; i.e., find
$\hat{\bm{\Theta}}^{(t+1)}= \arg\max_{\bm{\Theta}} ~ f_t(\bm{\Theta}).$
\myItemizeEnd
Note that the actual computation in the E-step is to generate sufficient statistics for computing $\arg\max_{\bm{\Theta}} \, f_t(\bm{\Theta})$, so that we do not need to scan the raw data every time when we need to evaluate $f_t(\bm{\Theta})$.  At the end, we obtain the MLE of $\bm{\Theta}$ modulo local maximums and Monte Carlo errors. We can then use the estimated $\hat{\bm{\Theta}}$ to obtain the posterior mean of the factors $(\bm{\eta} \,|\, \bm{y}, \hat{\bm{\Theta}})$ again through Gibbs sampling.  See~\cite{rlfm:kdd09} for an example of such an MCEM algorithm.

\parahead{Computational complexity}
We use a Gibbs sampler in the E-step, which is actually highly parallelizable. Take user and item factors for example.  Conditional on global factors $\bm{u}_i,\bm{v}_{j}$, the local factors $\bm{u}_{ik},\bm{v}_{jk}$ for
each action type can be sampled in parallel since they are only connected to each other through the global factors. When sampling local factors for each  action type,
we note that given $v_{jk}$s, the $u_{ik}$s are conditionally independent and can be sampled in parallel (similar assertion holds for $v_{jk}$s). Conditional on local factors, the global factors can also be sampled efficiently since the $\bm{u}_i$s and $\bm{v}_j$s are conditionally independent.  The complexity of
sampling a factor vector is at most $O(F^{3})$ and since $F$ is typical small, the E-step is computationally efficient. The major computation in the
M-step involves fitting standard linear regressions, which can also be parallelized. Thus, our MCEM algorithm is computationally efficient.
 We note that this training algorithm is similar to~\cite{reputation:kdd11,rlfm:kdd09} and logistic response can be handled by variational approximation~\cite{Jaakkola00}.  Thus, we omit the details and will provide links to our code and detailed derivations.

\section{Experiments}
\label{sec:experiment}

We evaluate the models presented in Section~\ref{sec:model} using
post-read data collected from a major online new site.  We
collected post-read actions from 13,739 users, each of whom has at
least 5 actions for at least one facet, to 8,069 items, each of
which received at least one post-read action for each type.  As a
result, we obtain 2,548,111 post-read action events, where each
{\em event} is identified by (user, facet, item). If the user took
an action on the item in the facet, the event is positive or {\em
relevant} (meaning that the item is relevant to the user in the
facet); if the user saw the item but did not take an action in the
facet, the event is negative or {\em irrelevant}. In this setting,
it is natural to treat each (user, facet) pair as a {\em query};
the set of events associated with that pair defines the set of items to be
ranked with relevance judgments coming from user actions.  Notice
that it is difficult to use editorial judgments in our setting
since different users have different preferences for their news
consumption.

\parahead{Evaluation metrics}
We use mean precision at $k$ (P@k) and mean average precision (MAP)
as our evaluation metrics, where mean is taken over the test (user,
facet) pairs. P@k of a model is computed as follows: For each test
(user, facet) pair, we use the model predictions to rank the items
seen by the user in that facet and compute the precision at rank
$k$, and then average the precision numbers over all the test
pairs. MAP is computed in the similar way.  To help comparison
among different models, we define {\em P@k Lift} and {\em MAP Lift}
over SMF of a model as the lift in P@k and MAP of the model over
the SMF model, which is a strong baseline defined in
Section~\ref{sec:baseline-model}.  Take P@k for example; if P@k of
a model is $A$ and P@k of SMF is $B$, then the lift is
$\frac{A-B}{B}$.

\parahead{Experimental setup} We create a training set, a tuning set and
a test set as follows. For each
user, we randomly select one facet in which he/she took some action
and put the events associated with this (user, facet) pair into set
$\mathcal{A}$ (for tuning and testing).
The rest of the (user, facet) pairs form the {\em
training set}.
Recall that each (user, facet) pair represents a query as in
standard retrieval tasks.
We then put 1/3 of set $\mathcal{A}$ into the {\em tuning
set} and the rest 2/3 into the {\em test set}.  The tuning set is
used to select the number of latent dimensions of the factor models
(i.e., the numbers of dimensions of $\bm{u}_i, \bm{v}_j, \bm{w}_k,
\bm{u}_{ik}, \bm{v}_{jk}$).  Notice that the EM-algorithm used in
our paper automatically determines all the model parameters except
for the number of latent dimensions. For each model, we only report
the test-set performance of the best number of dimensions selected
using the tuning set.

The user features used in our experiments consist of age, gender
and geo-location identified based on users' IP addresses.  We only
consider logged-in users; their user IDs are anonymized and not
used in any way. The item features consists of article categories
tagged by the publishers and the bag of words in the article titles
and abstracts.

\parahead{Models}
We compare the following models:
\myItemizeBegin
\item {\em LAT:} Locally augmented tensor model (Section~\ref{sec:lat-model}).
\item {\em BST:} Bias-smoothed tensor model, which is a special case of LAT (Section~\ref{sec:lat-model}).
\item {\em SMF:} Separate matrix factorization (Section~\ref{sec:baseline-model}).
\item {\em CMF:} Collapsed matrix factorization (Section~\ref{sec:baseline-model}).
\item {\em Bilinear:} This model uses the user features $\bm{x}_i$ and item features $\bm{x}_j$ to predict whether a user would take an action on an item. Specifically,
$$
y_{ijk} \sim \bm{x}'_i W_k \bm{x}_j,
$$
where $W_k$ is the regression coefficient matrix for facet $k$.
Notice that this model has a regression coefficient for every pair
of an individual user feature and an individual item feature, which
is fitted using Liblinear~\cite{Liblinear} with $L_2$
regularization, where the regularization weight is selected using
5-fold cross-validation.
\myItemizeEnd
We also compare the above models to a set of baseline IR models. In
all of the following IR models, we build a user profile based on the
training data by aggregating all the text information of the items
on which the user took positive actions. We treat such user profiles
as queries and then use different retrieval functions to rank the
items. The IR models include:
\myItemizeBegin
\item {\em COS:} Vector space model with cosine similarity.
\item {\em LM:} The Dirichlet smoothed language model~\cite{Zhai:01a}.
\item {\em BM25:} The best variant of Okapi retrieval methods~\cite{okapi}.
\myItemizeEnd
For the factor models, we note that the Gaussian version gives
better MAP values on the tuning set than the logistic version; so,
we report the performance of the Gaussian version.

\begin{figure} \centering
\epsfig{file=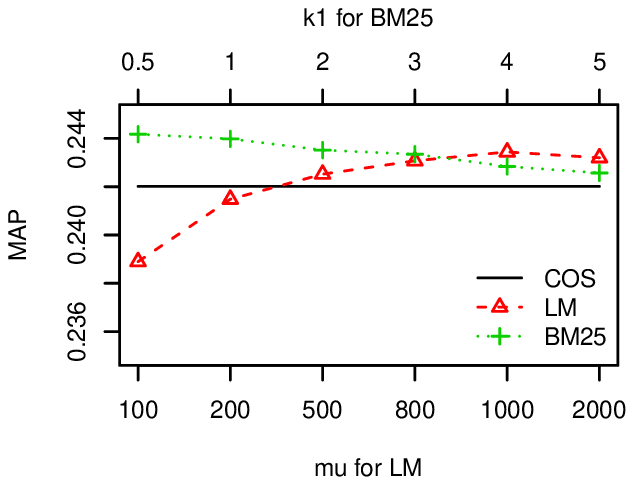, width=2.2in}
\vspace{-0.1in}
\caption{Performance of different IR models}
\label{fig:ir-model}
\vspace{-0.2in}
\end{figure}

\parahead{Performance of IR Models}
We first compare the baseline IR models in
Figure~\ref{fig:ir-model}.  In this figure, we vary parameter
$\mu$ for LM and parameter $k_1$ for BM25. The other two
parameters are set at the recommended default values $k_3=1000$ and
$b=0.75$ in all the experiments. From this figure, we can see that
both LM and BM25 can outperform COS, but the
difference is not large. In the following, we use the BM25
with $k_1=1$ as the IR model to compare with other learning-based
methods.

\begin{figure*}
\centering
\vspace{-0.25in}
\hspace{-0.4in}
\subfigure[Precision-recall curves]{\label{fig:P-R-curves}
	\epsfig{file=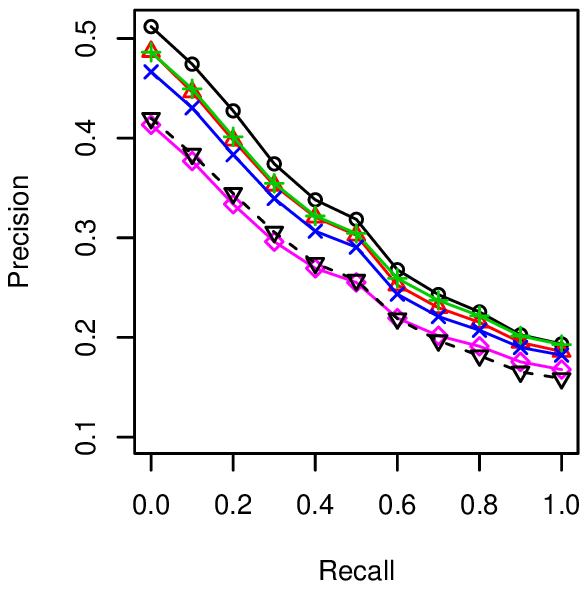, width=2.2in, angle=270}
}
\subfigure[P@1 by user activity levels]{\label{fig:P1-by-users}
	\epsfig{file=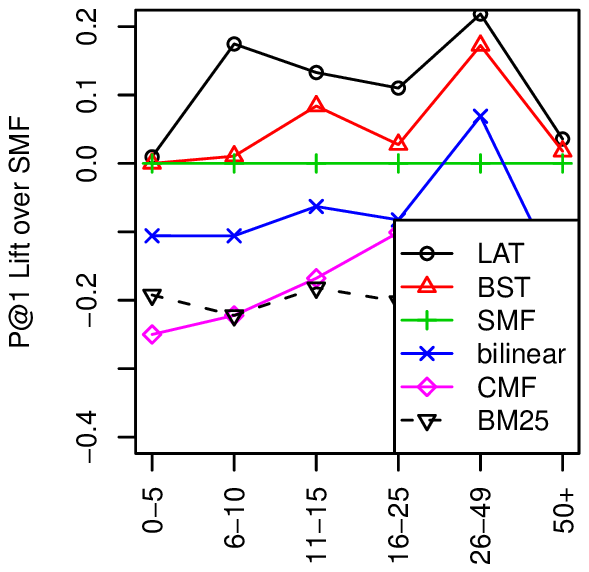, width=2.2in, angle=270}
}
\subfigure[MAP by user activity levels]{\label{fig:MAP-by-users}
	\epsfig{file=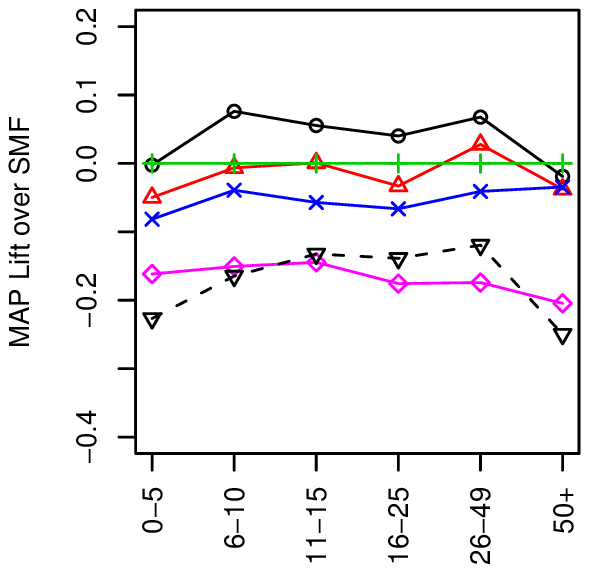, width=2.2in, angle=270}
}
\hspace{-0.4in}
\vspace{-0.1in}
\caption{Performance of different models}
\label{fig:rating-pred}
\vspace{-0.05in}
\end{figure*}

\parahead{Overall performance}
The precision-recall curves averaged over all (user,facet) pairs in
the test data of different models are shown in
Figure~\ref{fig:P-R-curves}, and P@1, P@3, P@5 and MAP are reported
in Table~\ref{tbl:overall-prec}. Notice that as $k$ increases, the
precision drops.  It is because post-read actions are rare events;
many users do not have 3 or 5 post-read actions in the test set.
For example, if a user only had one action and saw at least five
items in the test set, his/her P@5 is at most 1/5.  To test the
significance of the performance difference between two models, we
look at P@k and MAP for each individual (user, facet) pair and
conduct paired t-test for the two models over all test (user,
facet) pairs.  The test result is shown in Table~\ref{tbl:t-test}.
In particular, LAT significantly outperforms all other models. We
find that the difference between BST and SMF and the difference
between CMF and BM25 are insignificant.

We defer the comparison between LAT, BST and SMF to the breakdown
analysis below.  Here, we note that Bilinear outperforms CMF because
CMF completely ignores the behavioral differences among action types.
The fact that Bilinear outperforms CMF shows that user and item
features have some predictive power, but compared to SMF, these
features are not sufficient to capture the behavior of individual
users or items.  We also note that BM25 is one of the worst performing models
probably because it is the only model without supervised learning.

\begin{table}
\begin{center}
\caption{Overall performance of different models}
\label{tbl:overall-prec}
\begin{tabular}{|r|rrrr|}\hline
 & \multicolumn{4}{|c|}{\bf Precision} \\
{\bf Model} & P@1 & P@3 & P@5 & MAP\\
\hline
LAT      & {\bf 0.3180} & {\bf 0.2853} & {\bf 0.2648} & {\bf 0.3048}\\
BST      & 0.2962 & 0.2654 & 0.2486 & 0.2873\\
SMF      & 0.2827 & 0.2639 & 0.2469 & 0.2910\\
Bilinear & 0.2609 & 0.2472 & 0.2350 & 0.2755\\
CMF      & 0.2301 & 0.2101 & 0.2005 & 0.2439\\
BM25     & 0.2256 & 0.2247 & 0.2207 & 0.2440\\
\hline
\end{tabular}
\vspace{-0.2in}
\end{center}
\end{table}

\begin{table}
\begin{center}
{\small
\caption{Paired t-test result.  Note that smaller level values represent stronger significance.}
\label{tbl:t-test}
\begin{tabular}{|r@{~}c@{~}l|l|}\hline
\multicolumn{3}{|c|}{\bf Comparison} & {\bf Significance level} \\
\hline
LAT &>& BST  & 0.05 (P@1), ~$10^{-4}$ (P@3, P@5, MAP)\\
     && Rest & $10^{-4}$ (all metrics) \\
\hline
BST &$\approx$& SMF & insignificant \\
\hline
BST &>& Bilinear & $10^{-3}$ (all metrics) \\
\hline
SMF &>& Bilinear & 0.05 (P@1), ~$10^{-3}$ (P@3, P@5, MAP) \\
\hline
BST &>& CMF  & $10^{-4}$ (all metrics) \\
SMF & & BM25 & \\
\hline
Bilinear &>& CMF  & $10^{-3}$ (all metrics) \\
         & & BM25 & \\
\hline
CMF &$\approx$& BM25 & insignificant\\
\hline
\end{tabular}
\vspace{-0.2in}
}
\end{center}
\end{table}

\parahead{Breakdown by facets}
In Table~\ref{tbl:P1-by-facet}, we break the test data down by facets and report P@1 for different models; the results for other metrics are similar.  Here, we focus on the comparison between LAT, BST and SMF.  Starting with BST vs. SMF, we see that BST outperforms SMF for the first three facets but underperforms for the last two facets.  We note that the first three facets have more events in our dataset than the last two.  The advantage of BST over SMF is that it has global factors; thus, the training actions in one facet are utilized to predict the test actions in other facets through the correlation among facets.  However, BST is less flexible than SMF.  In particular, it is not flexible enough to capture the differences among facets; thus, it is forced to fit some facets better than others.  As expected, it fits the actions in facets with more data better than those with less data.  LAT addresses this problem by adding facet-specific factors ($\bm{u}_{ik}$ and $\bm{v}_{jk}$) to model the residuals of BST.  As can be seen, LAT uniformly outperforms BST.  It also outperforms SMF except for Mail.  The fact that SMF and Bilinear have the same performance for Mail suggests the difficulty of using latent factors to improve accuracy.  Since LAT has more factors than SMF, it has a higher chance of overfitting.

\begin{table}
\begin{center}
{\small
\caption{P@1 broken down by Facets}
\label{tbl:P1-by-facet}
\vspace{-0.1in}
\begin{tabular}{|r|rrrrr|}\hline
 & \multicolumn{5}{|c|}{\bf Facet} \\
{\bf Model} & Comment & Thumb & Facebook & Mail & Print \\
\hline
LAT      & {\bf 0.3477} & {\bf 0.3966} & {\bf 0.2565} &  0.2069 & {\bf 0.2722}\\
BST      & 0.3310 & 0.3743 & 0.2457 &  0.1936 & 0.1772\\
SMF      & 0.2949 & 0.3408 & 0.2306 &  {\bf 0.2255} & 0.2532\\
Bilinear & 0.2837 & 0.2947 & 0.2328 &  {\bf 0.2255} & 0.1709\\
CMF      & 0.2990 & 0.2905 & 0.1638 &  0.1114 & 0.1203\\
BM25     & 0.2726 & 0.3198 & 0.1509 &  0.1061 & 0.0886\\
\hline
\end{tabular}
}
\vspace{-0.2in}
\end{center}
\end{table}

\commentout{
\begin{table}
\begin{center}
{\small
\begin{tabular}{|r|rrrrr|}\hline
 & \multicolumn{5}{|c|}{\bf Action Types} \\
{\bf Model} & Comment & Thumb & Facebook & Mail & Print \\
\hline
LAT      & {\bf 1.724} & {\bf 1.766} & {\bf 2.363} &  2.546 & {\bf 2.494}\\
BST      & 1.593 & 1.610 & 2.222 &  2.318 & 1.275\\
SMF      & 1.310 & 1.377 & 2.024 &  {\bf 2.864} & 2.250\\
Bilinear & 1.223 & 1.055 & 2.053 &  {\bf 2.864} & 1.194\\
CMF      & 1.343 & 1.026 & 1.148 &  0.909 & 0.544\\
Cosine   & 0.765 & 0.977 & 0.215 & -0.090 & 0.463\\
\hline
\end{tabular}
\caption{Top-$1$ precision lift over the random model broken down by action types}
\label{tbl:prec-by-type}
}
\vspace{-0.15in}
\end{center}
\end{table}
}

\parahead{Breakdown by user activity levels}
In Figure~\ref{fig:P1-by-users} and \ref{fig:MAP-by-users}, we break test users down by their activity levels in terms of the numbers of post-read actions that they took in the training data.  Here, our focus is also on comparing LAT and BST to SMP.  Each curve represent the P@1 Lift or MAP Lift of each model over the SMF model as a function of the user activity level specified on the x-axis.  As can be seen, LAT almost uniformly outperforms all other models.  For users with low activity levels (0-5), there is almost no difference between LAT, BST and SMF because they all lack data and the predictions are mostly based on features.  For users who took 5-50 post-read actions, we see the largest advantage of using LAT.

\commentout{
\begin{table}
\begin{center}
{\small
\begin{tabular}{|r|rrrrrr|}\hline
 & \multicolumn{6}{|c|}{\bf Number of Training Actions per User} \\
{\bf Model} & 0-5 & 6-10 & 11-15 & 16-25 & 26-49 & 50+ \\
\hline
LAT      & {\bf 1.498} & {\bf 2.125} & {\bf 2.218} & {\bf 1.871} & {\bf 1.818} & {\bf 1.768}\\
BST      & 1.474 & 1.688 & 2.079 & 1.658 & 1.712 & 1.720\\
SMF      & 1.474 & 1.660 & 1.841 & 1.587 & 1.313 & 1.672\\
bilinear & 1.212 & 1.379 & 1.662 & 1.373 & 1.472 & 1.195\\
CMF      & 0.855 & 1.069 & 1.364 & 1.326 & 0.967 & 1.100\\
cosine   & 0.760 & 0.985 & 0.808 & 0.305 & 0.568 & 0.241\\
\hline
\end{tabular}
\caption{Top-$1$ precision lift over the random model broken down by user activity levels over all types}
\label{tbl:prec-by-level}
\vspace{-0.15in}
}
\end{center}
\end{table}
}

\parahead{Perceived differences among facets}
In Table~\ref{tbl:casestudy}, we show some examples of the result of multi-faceted news ranking.  On the top half of the table, we
show top-ranked articles for an average user. On the bottom half,
we show top-ranked articles for males with ages between 41 and 45.
From this table, we can see that different facets have very
different ranking results. For example, in the
Facebook and Mail facets, many health-related articles are highly ranked.
But for the Comment facet, political articles are usually
preferred. Furthermore, if we compare the males in the middle age
with the overall population, we also see notable differences. For
example, although both populations have health-related articles in the Mail facet,
middle-age males tend to mail more cancer-related
articles. These differences confirm the need for personalized multi-facet ranking and our proposed method can
address this need in a principled way.

\begin{table*}\small
\hspace{-8mm}
\begin{tabular}{|l|l|l|}
\hline \multicolumn{1}{|c|}{\textbf{Facebook}} & \multicolumn{1}{|c|}{\textbf{Mail}} & \multicolumn{1}{|c|}{\textbf{Comment}} \\
\hline \multicolumn{3}{|c|}{\textbf{Overall population}} \\
\hline
US weather tornado Japan disaster aid &	Teething remedies pose fatal risk to infants & US books Michelle Obama \\
\hline
Eight ways monsanto is destroying our health & US med car seats children & US Obama immigration \\
\hline
Teething remedies pose fatal risk to infants & Super women mom soft wins may live longer & US exxon oil prices \\
\hline
New zombie ant fungi found & Tips for a successful open house & Harry Reid: republicans fear tea party \\
\hline
Indy voters would rather have Charlie Sheen ... & Painless diabetes monitor talks to smartphone & Obama to kick off campaign this week \\

\hline \multicolumn{3}{|c|}{\textbf{For male at age 41 to 45}} \\	
\hline
Oxford English dictionary added new words & Richer white women more prone to melanoma & Israel troubling tourism \\
\hline
US exxon oil prices & Obesity boost aggressive breast cancer in older women & Israel palestinians \\
\hline
Children make parents happy eventually & US med car seats children & USA election Obama \\
\hline
Qatar Saudi politics Internet & Are coffee drinkers less prone to breast cancer & US books Michelle Obama \\
\hline
Lawmakers seek to outlaw prank calls & Short course of hormone therapy boosts prostate cancer & Levi Johnston to write memoir \\
\hline
\end{tabular}
\caption{Examples of multi-faceted news ranking. Only the titles of
the news articles are shown.} \label{tbl:casestudy}
\end{table*}

\section{Related Work} \label{sec:related}



Algorithmic news recommendation has received considerable attention
recently. Traditional recommendation approaches include
content-based filtering and collaborative filtering
techniques~\cite{YiZhang:SIGIR07,CF:01,Yehuda:KDD08}. These
techniques have been successfully applied to applications like
movie or product recommendation~\cite{CF:01,Linden:03}. In
particular, matrix factorization based collaborative filtering,
which belongs to the family of latent factor models, have achieved
the state-of-the-art accuracy~\cite{Yehuda:KDD08}. Recently, these
methods have been adapted for news recommendation. For example,
\cite{Newsjunkie:04} studied information novelty using
content-based methods. In \cite{DasWWW07}, collaborative filtering
was leveraged in an online news recommendation system. Hybrid
approaches which combine both content-based and collaborative
methods are also studied in news recommendation
recently~\cite{SCENE:SIGIR11}. In the news domain, the existing
work mainly ranks articles using clicks as the metric.
Some recent work starts looking into other metrics such as social
sharing~\cite{Share:2011}. To the best of our knowledge, little
prior work has studied the news recommendation in a multi-faceted
view, which becomes natural along with the popularity of Web 2.0.
In our work, we define facets according to post-read actions and
provide detailed analysis which shows the importance of
multi-faceted news ranking. Furthermore, a novel matrix
factorization based method to jointly model multi-type post-read
actions is proposed.

Our work is related to faceted
search~\cite{HearstFacet:03,FacetYiZhang:SIGIR10,Dumais:09}. The
goal of faceted search is to use facet metadata of a domain to help
users narrow their search results along different dimensions. In
the most recent TREC Blog track 2009~\cite{blogtrack2009}, a
special track of ``faceted blog distillation'' is initiated and the
task of this track is to find results relevant to a single facet of
a query such as ``opinionated'' articles in the blog collection. In
these types of work, facets are metadata related to contents. The
facets in our definition are based on user post-read actions and
our multi-faceted ranking is to help users quickly get interesting
news articles according to their preferred actions. Thus our work
provides a novel angle to define facets.

The technique used in our paper is closely related to latent factor
models such as matrix factorization or tensor decomposition. For
example, singular value decomposition (SVD) based
methods~\cite{Yehuda:KDD08}, tensor-based
methods~\cite{Rendle:SIGIR11}, and collaborative competitive
filtering~\cite{CCF:SIGIR11} all belong to this family. All these
methods did not consider post-read actions. In particular, our
technique is related to the collective matrix factorization
(CMF)~\cite{Singh08relationallearning} and bias-smoothed tensor
(BST) model~\cite{reputation:kdd11}. As compared in our
experiments, our models are better than these existing ones in
exploring the correlations among different post-read facets.

\section{Conclusion}
\label{sec:conclusion}
Jointly mining and modeling post-read actions of multiple types has not been previously studied in the literature.
We conducted a rigorous study on post-read behavior on Yahoo! News with action types like facebook share, commenting, rating that users engage in after reading an article.
Through data analysis, we found some interesting patterns in news consumption when it comes to read and post-read behavior. Reading articles is private,
post-read behavior like sharing and commenting are more public. Users tend to differ in interesting ways in their public and private behavior when 
interacting with news. We also saw huge variation in post-read action rates at the article resolution relative to classical measures like click-rates that
are used in recommending articles, perhaps providing a plausible explanation of why such engagment metrics have not been incorporated into news recommendation
algorithms before. However, we found positive correlations among different action types and were able to exploit these through a novel factor model called
Locally Augmented Tensor (LAT) to improve predictive performance of post-read action rates. This opens the door to incorporate post-read engagement behavior
in creating new products/modules on news sites online. We plan to explore such possibilities in the future.

{\small
\bibliographystyle{abbrv}
\bibliography{paper}
}
\end{document}